# Correlation functions between monopoles and instantons *

Harald Markum, Wolfgang Sakuler and Stefan Thurner

Institut für Kernphysik, Technische Universität Wien, A–1040 Vienna, Austria

We analyze the relation between instantons and abelian projected monopoles in both phases of pure QCD by calculating local correlation functions between topological charge densities and monopole densities. On an $8^3 \times 4$ lattice, it turns out that topological quantities are correlated approximately two lattice spacings. The monopole-instanton correlations are rather insensitive under cooling of gauge fields.

*1. Introduction.* It is assumed that gauge field configurations with non-trivial topology can explain the low-energy properties of QCD and thus could provide a solution to the confinement problem. There are two different kinds of topological objects which seem to be important for the confinement mechanism: color magnetic monopoles and instantons. Color magnetic monopoles play the essential role in the dual superconductor hypothesis [1]. In this picture confinement emerges by condensation of abelian monopoles via the dual Meissner effect. There is strong evidence from lattice calculations that the idea of dual superconductivity is an adequate description of the confinement phenomenon. On the other hand the role of instantons with respect to confinement is not so clear. It is suggested that instantons can only cause confinement in QCD by forming a so-called instanton liquid [2]. A fascinating question is whether color magnetic monopoles and instantons are related or not.

In a recent study we analyzed the distribution of the topological charge density and the monopole density around static color sources [3]. We observed that both topological quantities show qualitatively the same behavior, namely a suppression in the vicinity of the external sources. Motivated by this similarity we calculate local correlation functions between the monopole density and the topological charge density [4]. Results for the size of the correlations between topological objects are reported.

*Supported in part by "Fonds zur Förderung der wissenschaftlichen Forschung" under Contract P9428-PHY.

*2. Theory.* In order to investigate monopole currents one has to project $SU(3)$ onto its abelian degrees of freedom, such that an abelian $U(1) \times U(1)$ theory remains [5]. This aim can be achieved by various gauge fixing procedures. We employ the so-called maximal abelian gauge which is most favorable for our purposes. After gauge fixing pure QCD may be regarded as a theory of color charges and color magnetic monopoles. A gauge transformation of a gauge field element $U(x,\mu)$ is given by

$$\tilde{U}(x,\mu) = g(x)U(x,\mu)g^\dagger(x+\hat{\mu}) \ , \qquad (1)$$

where $g(x) \in SU(3)$. The maximal abelian gauge is imposed by maximizing the functional

$$R = \sum_{x,\mu,i} |\tilde{U}_{ii}(x,\mu)|^2. \qquad (2)$$

To extract abelian parallel transporters one has to perform the decomposition

$$\begin{aligned} \tilde{U}(x,\mu) &= c(x,\mu)u(x,\mu) \ , \quad \text{with} \qquad (3) \\ u(x,\mu) &= \text{diag}\,[u_1(x,\mu), u_2(x,\mu), u_3(x,\mu)] \ , \\ u_i(x,\mu) &= \exp\left[i \arg \tilde{U}_{ii}(x,\mu) - \frac{1}{3} i\phi(x,\mu)\right] \ , \\ \phi(x,\mu) &= \sum_i \arg \tilde{U}_{ii}(x,\mu)\Big|_{\text{mod } 2\pi} \in (-\pi,\pi] \ . \end{aligned}$$

Since the maximal abelian subgroup $U(1) \times U(1)$ is compact, there exist topological excitations. These are color magnetic monopoles which have integer-valued magnetic currents on the links of the dual lattice:

$$m_i(x,\mu) = \frac{1}{2\pi} \sum_{\square \ni \partial f(x+\hat{\mu},\mu)} \arg u_i(\square) \ , \qquad (4)$$

where $u_i(\Box)$ denotes a product of abelian links $u_i(x,\mu)$ around a plaquette $\Box$ and $f(x+\hat{\mu},\mu)$ is an elementary cube perpendicular to the $\mu$ direction with origin $x+\hat{\mu}$. The magnetic currents form closed loops on the dual lattice as a consequence of monopole current conservation. Finally the local monopole density is given by

$$\rho(x) = \frac{1}{3\cdot 4V_4}\sum_{\mu,i} |m_i(x,\mu)| \ . \tag{5}$$

For the implementation of the topological charge on the lattice there exists no unique discretization. In this work we restrict ourselves to the so-called field theoretic definitions which approximate the topological charge in the continuum [6]

$$q(x) = \frac{g^2}{32\pi^2}\epsilon^{\mu\nu\rho\sigma}\ \mathrm{Tr}\ \Big(F_{\mu\nu}(x)F_{\rho\sigma}(x)\Big)\ , \tag{6}$$

in the following ways:

$$q^{(P,H)}(x) = -\frac{1}{2^4 32\pi^2}\sum_{\mu,\ldots=\pm 1}^{\pm 4} \tilde{\epsilon}_{\mu\nu\rho\sigma}\mathrm{Tr}\ O^{(P,H)}_{\mu\nu\rho\sigma}, \tag{7}$$

with

$$O^{(P)}_{\mu\nu\rho\sigma} = U_{\mu\nu}(x)U_{\rho\sigma}(x)\ , \tag{8}$$

for the plaquette prescription and

$$\begin{aligned} O^{(H)}_{\mu\nu\rho\sigma} &= U(x,\mu)U(x+\hat{\mu},\nu)U(x+\hat{\mu}+\hat{\nu},\rho) \\ &\times\ U(x+\hat{\mu}+\hat{\nu}+\hat{\rho},\sigma)U^\dagger(x+\hat{\nu}+\hat{\rho}+\hat{\sigma},\mu) \\ &\times\ U^\dagger(x+\hat{\rho}+\hat{\sigma},\nu)U^\dagger(x+\hat{\sigma},\rho)U^\dagger(x,\sigma), \end{aligned} \tag{9}$$

for the hypercube prescription. The lattice and continuum versions of the theory represent different renormalized quantum field theories, which differ from one another by finite, non-negligible renormalization factors [7]. A simple procedure that enables one to get rid of renormalization constants, while preserving physical information contained in lattice configurations, is the cooling method. In our investigation we have employed the so-called "Cabbibo–Marinari method".

To measure the correlations between topological quantities we calculate the functions

$$\begin{aligned} &\langle q(0)q(d)\rangle\ , &&\langle q^2(0)q^2(d)\rangle\ , \\ &\langle \rho(0)\rho(d)\rangle\ , &&\langle \rho(0)q^2(d)\rangle\ . \end{aligned} \tag{10}$$

Since topological objects with opposite sign are equally distributed, we correlate the monopole density with the square of the topological charge density.

*3. Results.* Our simulations were performed on an $8^3\times 4$ lattice with periodic boundary conditions using the Metropolis algorithm. The observables were studied in pure QCD with the plaquette action both in the confinement and deconfinement phase at inverse gluon coupling $\beta = 6/g^2 = 5.6$ and 5.8, respectively. We made 50000 iterations and measurements were taken after every 50th iteration. Each of these 1000 configurations was first cooled and then subjected to 300 gauge fixing steps enforcing the maximal abelian gauge.

The correlation functions between topological quantities according to Eq. (10) are shown in Fig. 1 for several cooling steps at $\beta = 5.6$. They are normalized after the subtraction of their cluster values. The hypercube definition is used for $q$. The range of the instanton auto-correlations $qq$ and $q^2q^2$ which are originally $\delta$-peaked grows rapidly with cooling reflecting the occurance of extended instantons [8]. In contrast the $\rho\rho$-correlation decreases since monopole loops become dilute with cooling. The $\rho q^2$-correlation seems rather insensitive to cooling and clearly extends over more than two lattice spacings, indicating some non-trivial local correlation between monopoles and topological charges.

Fig. 2 presents row $\rho q^2$-correlations for the plaquette and the hypercube definition of the topological charge density in the confinement (l.h.s.) and the deconfinement phase (r.h.s.) for 6 and 11 cooling steps. The agreement between the two definitions is almost perfect. The absolute value of the correlations is smaller in the deconfinement phase since both instantons and monopoles become dilute. But there are practically no differences between the $\rho q^2$-correlations in the two phases after normalization (not shown). This holds also for the other topological correlation functions.

In Fig. 3 the sum of the squared topological charge density $K := \sum_x q^2(x)$ (hypercube definition) is plotted against the number of monopoles $N_{mon} = \sum_x \rho(x)$ for each configuration of our sample in the confinement phase. One observes that the slopes of the elliptic regions ascent with increasing cooling steps. This again indicates that monopoles and instantons are related.

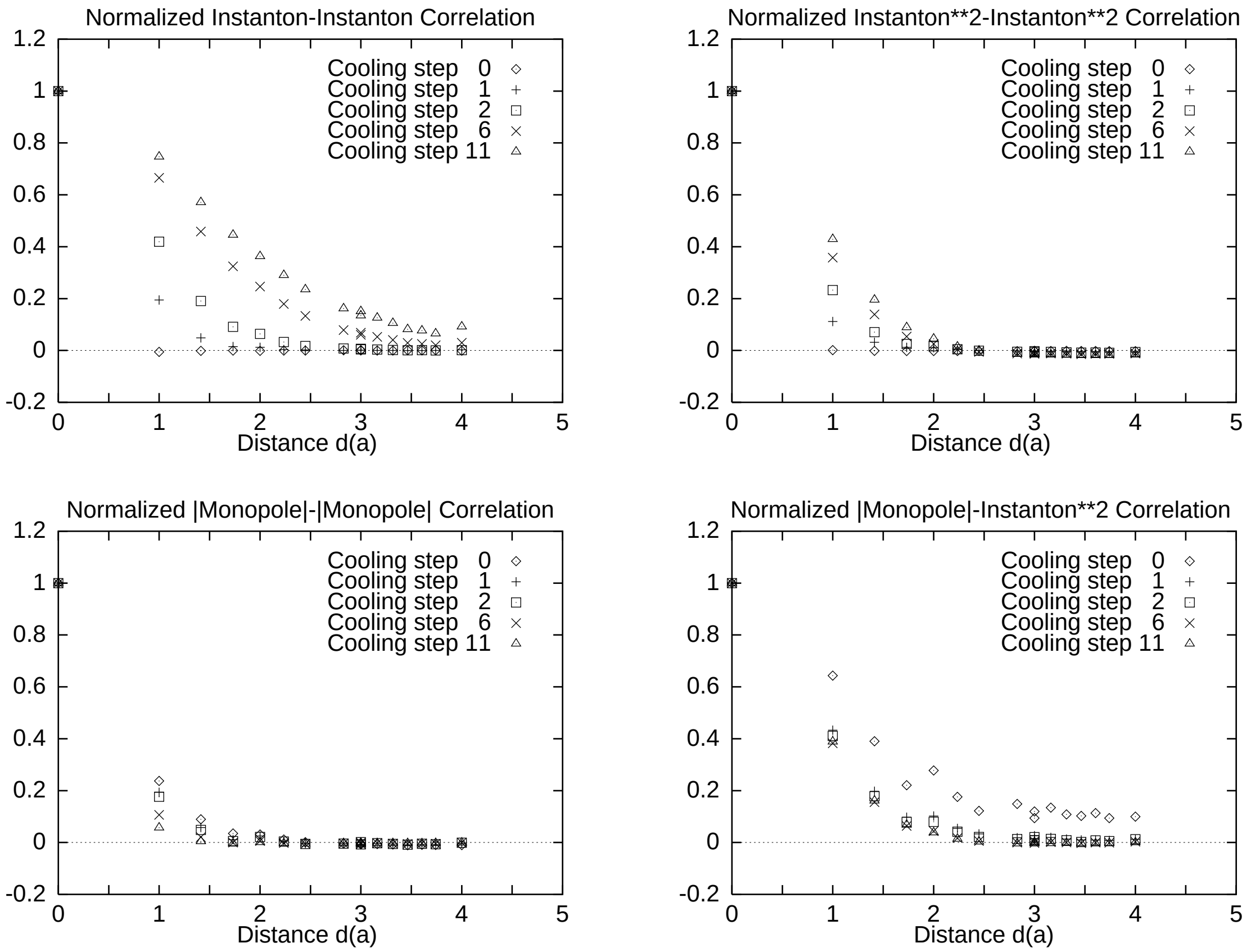

Figure 1. Correlation functions between topological charge densities and monopole densities in the confinement phase for $0, 1, 2, 6, 11$ cooling steps. The instanton auto-correlations grow with cooling reflecting the existence of extended instantons whereas the monopole auto-correlations decrease since the monopoles become diluted. The correlations between monopoles and instantons are nearly invariant under cooling with a range of approximately two lattice spacings indicating a non-trivial relation between these topological objects. Error bars are in the size of the anisotropy effects.

*4. Discussion and Outlook.* Calculations of correlation functions between monopole and topological charge densities yield a range of about two lattice spacings. Our results are in agreement with similar studies in $SU(2)$ [9] and provide indications for a non-trivial relation between monopoles and instantons. There is an enhanced probability that at the locations of instantons also monopoles can be found on gauge average. To get a deeper understanding of this relationship we analyze at present the correlations between color magnetic monopoles and instantons per gauge field configuration in more detail. Effects of sea quarks will be studied. Besides we plan to use a geometric definition of the topological charge for future investigations.

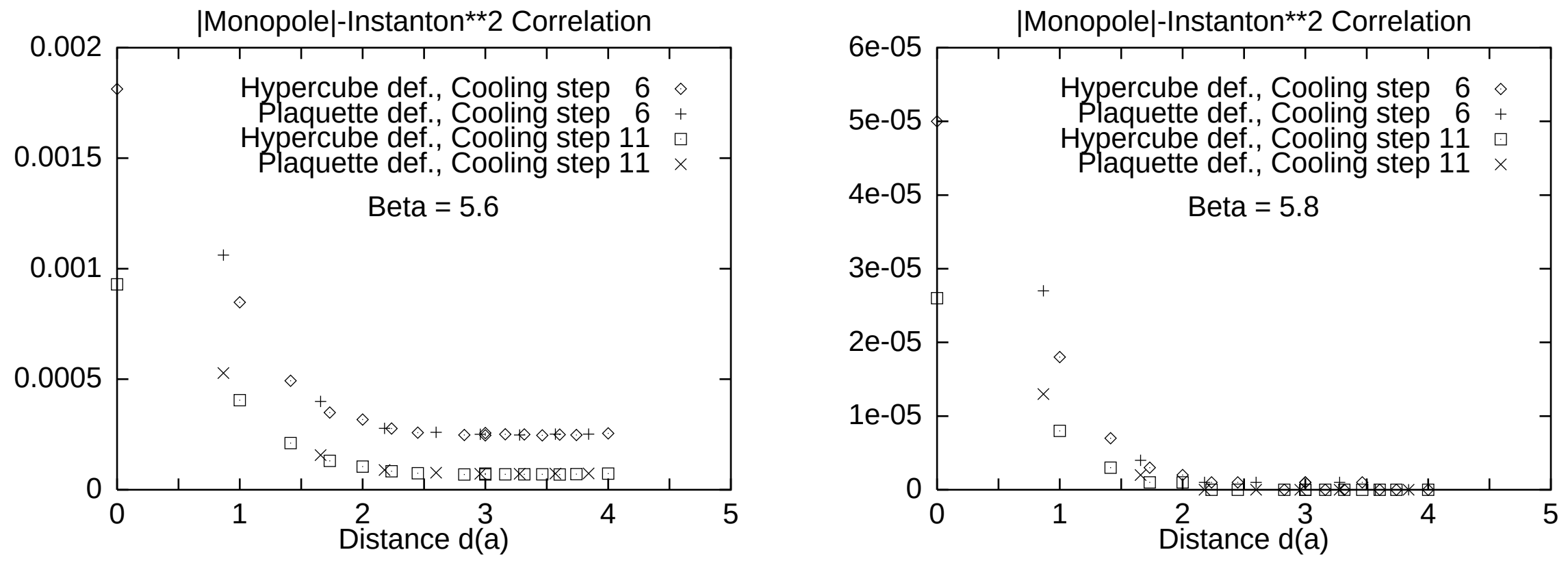

Figure 2. Correlation functions between the monopole density and the topological charge density for both definitions of $q$ in the confinement (l.h.s.) and the deconfinement phase (r.h.s.) for 6 and 11 cooling steps. The agreement between the plaquette and the hypercube definition is perfect. The row correlations are smaller in the deconfinement phase since both instantons and monopoles become dilute.

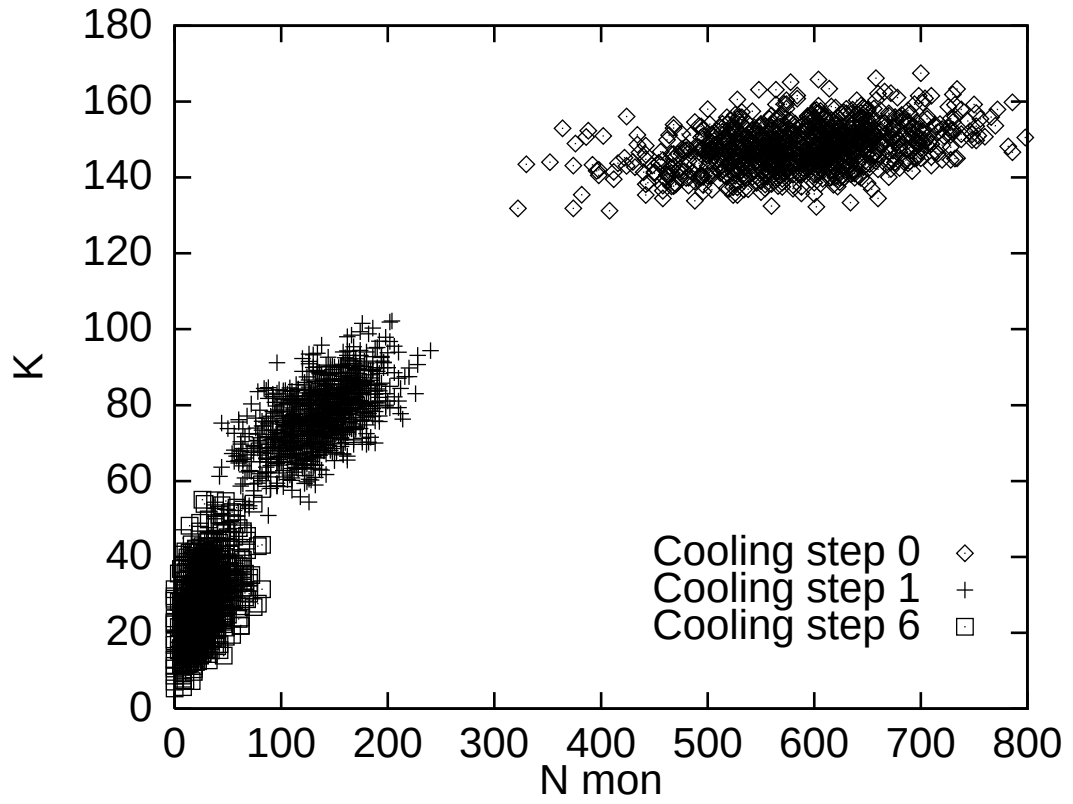

Figure 3. $K = \sum_x q^2(x)$ versus monopole number $N_{mon}$ for 1000 configurations. The slopes of the elliptic regions grow with cooling hinting at non-trivial instanton-monopole correlations.